\documentclass[useAMS,usenatbib]{mn2e}

\usepackage{natbib}
\usepackage{graphicx}
\usepackage{amssymb}
\usepackage{color}
\usepackage[dvipsnames]{xcolor}
\usepackage{caption}
\usepackage{lipsum,graphicx,multicol}
\usepackage{float}
\usepackage[fleqn]{amsmath}
\usepackage{subcaption}

\title[BH-Exoplanet link]{ How black hole activity may influence exoplanetary evolution in our Galaxy } 
\author[ ]
{W. Ishibashi$^{1}$\thanks{E-mail: wako.ishibashi@physik.uzh.ch} 
\footnotemark[0]\\
\footnotemark[0]\\
$^{1}$Physik-Institut, Universit$\ddot{a}$t Zurich, Winterthurerstrasse 190, 8057 Zurich, Switzerland \\
$^{2}$Istituto Ricerche Solari (IRSOL), Universit$\grave{a}$ della Svizzera italiana (USI), 6605 Locarno Monti, Switzerland \\
$^{3}$Euler Institute, Universit$\grave{a}$ della Svizzera italiana (USI), 6900 Lugano, Switzerland 
}

\begin{document}

\pdfminorversion=4

\date{Accepted 2024 July 22. Received 2024 July 18; in original form 2024 June 18 }

\pagerange{\pageref{firstpage}--\pageref{lastpage}} \pubyear{2012}

\maketitle

\label{firstpage}

\vspace{2cm}

\begin{abstract}
An increasing number of exoplanets have been discovered in the Milky Way galaxy, which is also known to harbour a super-massive black hole (Sagittarius A*) at its centre. Here, we investigate how the central black hole (BH) activity may affect the evolution of exoplanets in our Galaxy. Accreting BHs emit high-energy radiation --extreme ultraviolet and X-rays-- which can lead to XUV photoevaporation of the planetary atmospheres. We evaluate the atmospheric mass-loss using both theoretical estimates of the BH radiative output and observational constraints on the past activity history of Sgr A*. The resulting mass-loss is analysed as a function of the galactocentric distance. For the first time, we compute the exoplanet atmospheric evolution under BH irradiation by explicitly including the temporal evolution of the central luminosity output (i.e. the BH activity history). We obtain that Sgr A* could have a major impact on exoplanets located in the inner region of the Galaxy (e.g. Galactic bulge): a significant fraction of the atmospheric mass can be removed by BH irradiation; and in extreme cases, the initial atmosphere may be completely stripped away. Such mass-loss can have important consequences on the atmospheric chemistry and potential biological evolution. We discuss the physical implications for planetary habitability, and we also briefly consider the case of stellar-mass BHs. Overall, accreting black holes may play a significant role in the evolution of exoplanets in our Galaxy across cosmic time. 
\end{abstract}



\section{Introduction}

An ever increasing number of extrasolar planets have been discovered in the Milky Way galaxy, with more than 5500 exoplanets confirmed to date\footnote{https://exoplanetarchive.ipac.caltech.edu}. The atmospheric characterisation of planets beyond our solar system is becoming possible with the James Webb Space Telescope (JWST) \citep[e.g.][and references therein]{Kempton_Knutson_2024}. At the other end of the galactic scale, we now know that a super-massive black hole (SMBH) resides at the centre of our Galaxy. The presence of the central SMBH, known as Sagittarius A* (Sgr A*), has been firmly confirmed by infrared observations of stellar orbits \citep{Genzel_et_2024} and the radio images obtained by the Event Horizon Telescope \citep{EHT_2022_I}. 

In light of these two empirical facts, one may ask: what is the possible impact of the central SMBH on the evolution of exoplanets in the Milky Way galaxy? In particular, how the high-energy radiation from the accreting BH may affect the planetary atmospheres? This will have implications for the evolution of exoplanets in our Galaxy and their potential habitability over cosmic time. 

Following a mass accretion event, the central BH emits a huge amount of radiation into the surroundings, mostly at ultraviolet (UV) and X-ray wavelengths. As a result, rapidly accreting SMBH can be observed as bright active galactic nuclei (AGN) in distant galaxies --such as quasars and Seyfert galaxies; while lower luminosity AGNs are observed in the local Universe \citep[e.g.][and references therein]{Padovani_et_2017, Hickox_Alexander_2018}. At the present time, the central BH in the Milky Way galaxy is extremely quiescent, accreting far below its Eddington limit \citep{EHT_2022_V}. However, there are a number of observational indications suggesting that Sgr A* underwent much more active episodes in the past, e.g. as evidenced by the discovery of the Fermi and eROSITA bubbles \citep{Su_et_2010, Predehl_et_2020}. Indeed, the detection of these large bipolar gamma-ray and X-ray structures hints at a past energetic outburst from the Galactic Centre \citep{Yang_et_2022}. 

In the context of star-planet interaction, it is well known that the high-energy radiation (X-rays and extreme ultraviolet, or XUV radiation) from the host star can lead to photoevaporation of the exoplanet atmosphere and subsequent mass-loss. Such atmospheric escape, due to XUV photoevaporation, is likely responsible for shaping the exoplanet `radius gap', characterised by a dearth of planets around $\sim 1.8 R_{\oplus}$, separating smaller super-Earths and larger sub-Neptunes \citep[][and references therein]{Fulton_et_2017, Bean_et_2021}. This results in a bimodal radius distribution --the radius valley-- with rocky cores stripped of their gaseous envelopes on the one side, and gas-rich planets still partially keeping their atmospheres on the other side. A possibility is that XUV radiation from the host star induces a transition from gaseous sub-Neptunes into rocky super-Earths.  

While the effects of host star irradiation on planetary atmospheres have been extensively analysed in the literature, only a limited number of studies have examined the impact of radiation from the central SMBH \citep{Balbi_Tombesi_2017, Forbes_Loeb_2018, Chen_et_2018, Amaro-Seoane_Chen_2019, Lingam_et_2019, Wislocka_et_2019}. Here we further explore the potential effects of the SMBH on exoplanets in the Milky Way galaxy by considering different theoretical estimates of the XUV radiation from the accreting BH, as well as some observational constraints on the past activity history of Sgr A*. We estimate the resulting atmospheric mass-loss based on the energy-limited formalism, and analyse the total mass lost as a function of the galactocentric distance. For the first time, we directly combine the exoplanet atmospheric evolution with the BH activity history, by explicitly including the temporal evolution of the AGN luminosity output, $L(t)$. 

The paper is structured as follows. We first recall the basics of XUV-driven photoevaporation, and then estimate the XUV luminosity from simple power-law modelling of the AGN spectral energy distribution (Sect. \ref{Sect_XUV_photoevaporation}). We calculate the resulting atmospheric mass-loss in Sect. \ref{Sect_mass-loss_theoretical}, and compare with the mass-loss obtained from observational constraints on the past activity history of Sgr A* (Sect. \ref{Sect_mass-loss_observational}). In Sect. \ref{Sect_atmospheric_evolution}, we compute the exoplanet atmospheric evolution under SMBH irradiation for different AGN luminosity evolution histories and exoplanetary parameters. The analogous case of stellar-mass BHs is briefly treated in Sect. \ref{Sect_smBH}. We discuss the physical implications for planetary/galactic habitability, and outline some caveats alongside possible directions for future research (Sect. \ref{Sect_Discussion}). 


\section{XUV radiation-driven photoevaporation}
\label{Sect_XUV_photoevaporation}


\subsection{Energy-limited atmospheric escape}

Exoplanets subject to strong irradiation from the host star can undergo atmospheric mass-loss via hydrodynamic escape \citep[e.g. see  review by][and references therein]{Owen_2019}. XUV photons are absorbed in the upper region of the H/He atmosphere, where they dissociate and photoionise molecules and atoms; the resulting free electrons heat the ambient gas to high temperatures ($T \sim 10^4$ K). The weakly bound gas particles may then escape the gravitational potential of the planet, leading to atmospheric mass-loss. Such thermal escape is known as XUV-driven photoevaporation. 
 
A simple way to parametrise the resulting mass-loss is given by the energy-limited formalism \citep{Watson_et_1981}, which is commonly adopted in the literature:
\begin{equation}
\dot{M} = \frac{\epsilon \pi F_\mathrm{XUV} R_p^3}{G M_p} , 
\end{equation}
where $\epsilon$ is the evaporation efficiency, $F_\mathrm{XUV}$ is the XUV flux, and $M_p$ and $R_p$ are the planetary mass and radius, respectively. 
This can be re-expressed in terms of the planet bulk density $\rho_p = 3 M_p/4 \pi R_p^3$ and the XUV luminosity $L_\mathrm{XUV} = 4 \pi r^2 F_\mathrm{XUV}$ as
\begin{equation}
\dot{M} = \frac{3 \epsilon L_\mathrm{XUV}}{16 \pi G \rho_p r^2} , 
\label{Eq_Mdot_Lxuv}
\end{equation}
where $r$ is the distance from the radiation source. 

The evaporation efficiency, which parametrizes the fraction of the incident flux used for PdV expansion work, is typically of the order of ten percent ($\epsilon \sim 0.1$); although in reality the actual efficiency parameter depends on the planet properties and the incident XUV flux \citep{Owen_2019}. In the case of stellar irradiation, $L_\mathrm{XUV}$ is the XUV luminosity from the host star, which is roughly constant in the early saturated regime and subsequently decays as the stellar rotation rate decreases with age \citep[][and references therein]{Bonfanti_et_2021}. For our purposes, we need to estimate the XUV luminosity originating from accretion onto the SMBH at the centre of the Galaxy. 


\subsection{XUV radiation from accreting SMBH}

Based on the Soltan argument \citep{Soltan_1982}, most of the SMBH growth occurs through luminous accretion, observationally manifest as AGNs. Indeed, most accreting SMBH may spend a few percent of their lifetime as active galaxies, with a major fraction of the bolometric luminosity emitted at UV and X-ray wavelengths. 


\subsubsection{X-ray luminosity}

The X-ray spectra of AGNs can be described by power-law spectra, characterised by the photon index $\Gamma$
\begin{equation}
L_\mathrm{X} = \int_{\nu_\mathrm{i}}^{\nu_\mathrm{f}} C \nu^{-(\Gamma -1)} d \nu
= \frac{C}{2 -\Gamma} \left( \nu_\mathrm{f}^{2-\Gamma} - \nu_\mathrm{i}^{2-\Gamma} \right) ,
\label{Eq_LX}
\end{equation}
in the frequency range from $\nu_\mathrm{i} \sim 0.1$ keV to $\nu_\mathrm{f} \sim 100$ keV, and where $C$ is a normalisation constant \citep{Arcodia_et_2019}. The hard X-ray bolometric correction in the $(2-10)$ keV band is defined as $\kappa_\mathrm{X} = \frac{L_\mathrm{bol}}{L(2-10 \mathrm{keV})}$, where $\lambda_\mathrm{E} = L_\mathrm{bol}/L_\mathrm{E}$ is the Eddington ratio and $L_\mathrm{E}$ is the Eddington luminosity.
Thus the normalisation constant can be expressed as
\begin{equation}
C = \frac{\lambda_\mathrm{E} L_\mathrm{E}}{\kappa_\mathrm{X}} \frac{2 - \Gamma}{\nu_\mathrm{10keV}^{2-\Gamma} - \nu_\mathrm{2keV}^{2-\Gamma}} . 
\end{equation}

We note that both the X-ray photon index ($\Gamma$) and the X-ray bolometric correction ($\kappa_\mathrm{X}$) depend on the Eddington ratio, and may be expressed as a function of $\lambda_\mathrm{E}$.  
The photon index typically spans the range $\Gamma \sim 1.5-2.5$ in different AGN samples \citep{Just_et_2007, Shemmer_et_2008}. Moreover, the photon index correlates with the Eddington ratio, such that higher accretion rates lead to steeper X-ray spectra \citep{Ishibashi_Courvoisier_2010}. 

A best-fit relation of the form
\begin{equation}
\Gamma(\lambda_\mathrm{E}) = 0.27 \log \lambda_\mathrm{E} + 2.14 
\end{equation}
is obtained for a sample of accreting AGNs with reverberation mapping measurements, with a similar slope reported in other works \citep[][and references therein]{Liu_et_2021}. 
The X-ray bolometric correction is also correlated with the Eddington ratio, and the dependence can be fit by a relation of the form
\begin{equation}
\kappa_\mathrm{X}(\lambda_\mathrm{E}) = a \left[ 1 + \left( \frac{\lambda_\mathrm{E}}{b} \right)^c \right] 
\end{equation}
with $a = 7.51$, $b = 0.05$, $c = 0.61$, based on a large sample of AGNs covering several decades in luminosity \citep{Duras_et_2020}. 

Combining the above three relations, $\kappa_\mathrm{X}(\lambda_\mathrm{E})$, $\Gamma(\lambda_\mathrm{E})$, and $C(\lambda_\mathrm{E})$, we can calculate the X-ray luminosity (equation \ref{Eq_LX}) for different Eddington ratios. In the case of Sgr A*, the black hole mass is $M_\mathrm{BH} \sim 4 \times 10^6 M_{\odot}$, and the corresponding Eddington luminosity is $L_\mathrm{E} = 4 \pi G c m_p M_\mathrm{BH}/\sigma_T \sim 5 \times 10^{44}$ erg/s. We can now evaluate the X-ray luminosity, e.g. in the $(1-10)$ keV band, for different $\lambda_\mathrm{E}$. 
The resulting X-ray luminosity is $L_\mathrm{X} = 3.7 \times 10^{42}$ erg/s for $\lambda_\mathrm{E} = 0.1$, and $L_\mathrm{X} = 1.4  \times 10^{43}$ erg/s for $\lambda_\mathrm{E} = 1$. 


\subsubsection{EUV luminosity}

Similarly, the broad-band EUV continuum of AGNs may be described by a power-law 
\begin{equation}
L_\mathrm{EUV} = \int_{\nu_\mathrm{L}}^{\nu_\mathrm{max}} B \left( \frac{\nu}{\nu_0} \right)^{-\beta} d \nu 
\end{equation}
where $h \nu_{0} = 10$ eV, $h \nu_\mathrm{L} = 13.6$ eV, $h \nu_\mathrm{max} = 1$ keV, and $\beta = 1.5$ \citep{VandenBerk_et_2020, Inayoshi_et_2022}. 
The EUV normalisation constant ($B$) may be related to the X-ray normalisation constant ($C$) at 1 keV by  
\begin{equation}
B = L_\mathrm{\nu, 1keV} \left( \frac{\nu_\mathrm{1keV}}{\nu_0} \right)^{\beta} 
= C \nu_\mathrm{1keV}^{1-\Gamma} \left( \frac{\nu_\mathrm{1keV}}{\nu_0} \right)^{\beta}  . 
\end{equation}
The EUV luminosity can then be evaluated for different Eddington ratios. 
The resulting EUV luminosity of Sgr A* is $L_\mathrm{EUV} =  2.1  \times 10^{43}$ erg/s for $\lambda_\mathrm{E} = 0.1$, and $L_\mathrm{EUV} =  1.1 \times 10^{44}$ erg/s for $\lambda_\mathrm{E} = 1$. 


\subsubsection{XUV luminosity}

Finally, the total XUV luminosity ($L_\mathrm{XUV}$) is obtained by summing the X-ray and EUV luminosities, for a given Eddington ratio. As a result, we obtain $L_\mathrm{XUV} = 2.5 \times 10^{43}$ erg/s for $\lambda_\mathrm{E} = 0.1$; while for $\lambda_\mathrm{E} = 1$, we get $L_\mathrm{XUV} = 1.2 \times 10^{44}$ erg/s. We note that these values represent a non-negligible  fraction of the Eddington luminosity ($L_\mathrm{XUV}/L_\mathrm{E}  \sim 0.05-0.24$). The maximal upper limit on the XUV radiative output of Sgr A* is obtained by taking $L_\mathrm{XUV} \sim L_\mathrm{E} \sim 5 \times 10^{44}$ erg/s. A similar approach based on power-law modelling of the AGN broadband spectrum was also adopted in previous works \citep{Balbi_Tombesi_2017, Amaro-Seoane_Chen_2019}. These theoretically estimated XUV luminosities can now be used to analyse the potential impact of the SMBH radiation on exoplanets distributed throughout the Milky Way galaxy. 


\section{SMBH-driven atmospheric escape: theoretical estimates}
\label{Sect_mass-loss_theoretical}


\subsection{XUV irradiation from Sgr A*}

The XUV flux from Sgr A* can be estimated from the XUV luminosity (derived in Sect. \ref{Sect_XUV_photoevaporation})
\begin{equation}
F_\mathrm{XUV} = \frac{L_\mathrm{XUV}}{4 \pi r^2} , 
\end{equation}
where $r$ is the galactocentric distance. 

Table \ref{Table_flux} shows the unattenuated XUV flux from Sgr A* at different galactocentric radii and for different Eddington ratios. These values can be compared with the minimum XUV flux required for hydrodynamic wind launch $F_\mathrm{XUV, min} = 0.1 \, \mathrm{erg \, s^{-1} cm^{-2}}$ \citep{Bolmont_et_2017, Chen_et_2018, Wislocka_et_2019}, and the current solar XUV flux at Earth $F_\mathrm{XUV, \oplus} = 4.7 \, \mathrm{erg \, s^{-1} cm^{-2}}$ \citep{Forbes_Loeb_2018}. By comparison, the stellar flux from a solar-like host star is about $F_\mathrm{XUV, \star} = 5.6 \, \mathrm{erg \, s^{-1} cm^{-2}}$ at a distance of 1 AU, while it is $F_\mathrm{XUV, \star} = 850 \, \mathrm{erg \, s^{-1} cm^{-2}}$ at a separation of 0.1 AU \citep{Chen_et_2018}. 

From table \ref{Table_flux}, we see that the XUV flux from Sgr A* is greater than the minimum value ($F_\mathrm{XUV, SMBH} > F_\mathrm{XUV, min}$) in the Galactic bulge region (at $r \lesssim$ 1 kpc) for Eddington ratios $\lambda_\mathrm{E} \geq 0.1$; at higher accretion rates, this condition is satisfied up to a distance of a few kpc. Moreover, in the inner galactic region ($r \lesssim 100$ pc), the XUV flux from Sgr A* is higher than the current solar XUV flux at Earth. In fact, the XUV irradiation from the SMBH ($F_\mathrm{XUV, SMBH} \sim 10^4 \, \mathrm{erg \, s^{-1} cm^{-2}}$ at $r \sim 10$ pc) may exceed the stellar flux of a close-in planet ($F_\mathrm{XUV, \star} \sim 850 \, \mathrm{erg \, s^{-1} cm^{-2}}$ at 0.1 AU). Similar results, whereby the XUV flux from the central SMBH exceeds the host stellar flux, have also been reported in other studies \citep{Chen_et_2018, Wislocka_et_2019}. Therefore the high-energy radiation from the central SMBH can be an important contributor to the atmospheric evolution of exoplanets located in the inner region of the Galaxy. 

\begin{table}
\caption{
XUV flux from the central SMBH ($F_\mathrm{XUV, SMBH}$ in units of $[\mathrm{erg \, s^{-1} cm^{-2}}]$) for different accretion states of Sgr A* at different galactocentric radii. 
 }
\begin{center}
\begin{tabular}{|c|c|c|c|}
\hline 
  & $\lambda_\mathrm{E} = 0.1$ & $\lambda_\mathrm{E} = 1$ & $L_\mathrm{XUV} = L_\mathrm{E}$ \\
 \hline 
r = 10 pc & $2.2 \times 10^3$ & $1.1 \times 10^4$ & $4.5 \times 10^4$ \\
r = 100 pc & 22 & 106 & 451 \\
r = 1 kpc & 0.2 & 1.1 & 4.5 \\
\hline
\end{tabular}
\end{center}
\label{Table_flux}
\end{table}


\subsection{SMBH-induced atmospheric mass-loss}

We recall that in the energy-limited formalism, the mass-loss rate directly scales with the XUV flux and is inversely proportional to the planet bulk density (cf. equation \ref{Eq_Mdot_Lxuv})
\begin{equation}
\dot{M} = \frac{3 \epsilon L_\mathrm{XUV}}{16 \pi G \rho_p r^2} ,
\label{Eq_Mdot_Lxuv_bis}
\end{equation}
where a standard evaporation efficiency of $\epsilon = 0.1$ is assumed in the following. 

In table \ref{Table_Mdot}, we report the resulting mass-loss rates of a rocky exoplanet (with $\rho_p = 5 \, \mathrm{g cm^{-3}}$) at different galactocentric radii and for different Eddington ratios. We see that in the innermost region ($r \sim 10$ pc), the mass-loss rate spans the range $\dot{M} \sim (5 \times 10^8 - 10^{10})$ g/s; while at larger galactocentric radii ($r \sim 1$ kpc), the mass-loss rates are correspondingly lower, $\dot{M} \sim (5 \times 10^4 - 10^6)$ g/s. Overall, the mass-loss rate induced by the central SMBH can cover a wide range, from $\sim 5 \times 10^4$ g/s up to $\sim 10^{10}$ g/s, depending on the accretion state of Sgr A* and the exoplanet location within the Galaxy. Naturally, a gaseous planet with a lower bulk density (say $\rho_p \sim 1 \, \mathrm{g cm^{-3}}$) would undergo enhanced mass-loss rates. Empirical data suggest that gas giant planets in the Galactic bulge may lose a total mass greater than Earth's atmosphere during the active phase of Sgr A* \citep{Wislocka_et_2019}. 

\begin{table}
\caption{
Atmospheric mass-loss rate ($\dot{M}$ in units of [g/s]) of a rocky planet induced by XUV irradiation from Sgr A* in different accretion states and at different galactocentric radii. 
}
\begin{center}
\begin{tabular}{|c|c|c|c|}
\hline 
  & $\lambda_\mathrm{E} = 0.1$ & $\lambda_\mathrm{E} = 1$ & $L_\mathrm{XUV} = L_\mathrm{E}$ \\
 \hline 
r = 10 pc & $5.0 \times 10^8$ & $2.4 \times 10^9$ & $9.9 \times 10^9$ \\
r = 100 pc & $5.0 \times 10^6$ & $2.4 \times 10^7$ & $9.9 \times 10^7$ \\
r = 1 kpc & $5.0 \times 10^4$ & $2.4 \times 10^5$ & $9.9 \times 10^5$ \\
\hline
\end{tabular}
\end{center}
\label{Table_Mdot}
\end{table}

Integrating the mass-loss rate (equation \ref{Eq_Mdot_Lxuv_bis}), we obtain the total mass loss over time
\begin{equation}
\Delta M_{atm} = \frac{3 \epsilon L_\mathrm{XUV} \Delta t}{16 \pi G \rho_p r^2} , 
\end{equation}
where $\Delta t$ is the AGN episode duration. A characteristic AGN activity timescale is given by the Salpeter time, defined as $t_\mathrm{S} = \eta c \sigma_T / 4 \pi G m_p \sim 5 \times 10^7$yr, which is the time it takes to double the BH mass through Eddington-limited accretion with a canonical radiative efficiency of $\eta = 0.1$ \citep{Balbi_Tombesi_2017, Forbes_Loeb_2018}. 

\begin{figure}
\begin{center}
\includegraphics[angle=0,width=0.4\textwidth]{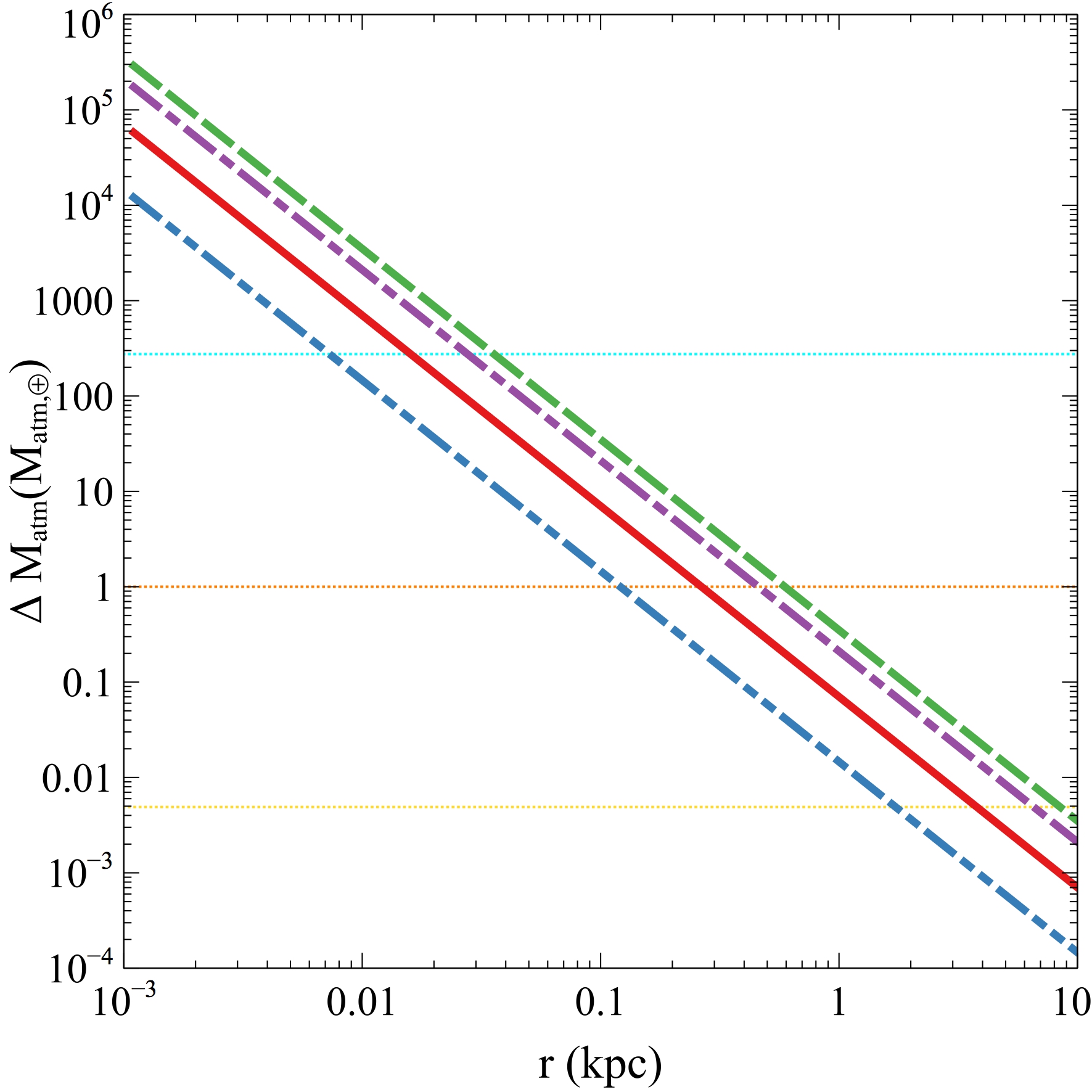} 
\caption{ 
Atmospheric mass-loss induced by the SMBH based on theoretical estimates of XUV irradiation. 
Fiducial model (red solid) and variations in SMBH and planetary parameters: $L_\mathrm{XUV} (\lambda_\mathrm{E} = 0.1) = 2.5 \times 10^{43}$ erg/s (blue dash-dot-dot), $\epsilon = 0.3$ (violet dash-dot), $\rho_p  = 1 \, \mathrm{g cm^{-3}}$ (green dashed). Horizontal lines: Earth's ocean mass ($\Delta M_\mathrm{atm}/M_\mathrm{atm,\oplus} = 275$, cyan fine dotted), Earth's atmospheric mass ($\Delta M_\mathrm{atm}/M_\mathrm{atm,\oplus} = 1$, orange fine dotted), and Mars atmospheric mass ($\Delta M_\mathrm{atm}/M_\mathrm{atm,\oplus} = 0.005$, yellow fine dotted). 
}
\label{Fig_DeltaM_r_model}
\end{center}
\end{figure}

In Fig. \ref{Fig_DeltaM_r_model}, we show the total mass loss $\Delta M_\mathrm{atm}$ (mass lost in units of the current Earth atmosphere mass $M_\mathrm{atm, \oplus}$) as a function of the galactocentric radius $r$. The three horizontal lines indicate the present-day Earth's ocean mass (cyan), the Earth's atmospheric mass (orange), and the Mars atmospheric mass (yellow), respectively. In the following, we assume as fiducial parameters of our model: $L_\mathrm{XUV} (\lambda_\mathrm{E} = 1) = 1.2 \times 10^{44}$ erg/s, $\epsilon = 0.1$, $\rho_p  = 5 \, \mathrm{g cm^{-3}}$, and $\Delta t = t_\mathrm{S} = 5 \times 10^7 $ yr (red solid line). For these fiducial values, we see that a mass equivalent to or greater than an Earth atmospheric mass can be lost at radii $r \lesssim 0.3$ kpc. Likewise, a mass equivalent to Mars atmospheric mass can be lost up to $r \sim 4$ kpc, while a mass comparable to Earth's ocean may be lost in the inner $\sim 10$ pc region. 

Variations in the SMBH and planetary parameters are also examined in Fig. \ref{Fig_DeltaM_r_model}. As expected, a lower XUV luminosity leads to lower mass-loss (blue line), but even with a reduced luminosity the central BH can still remove an Earth atmosphere mass up to $r \sim 0.1$ kpc. A lower bulk density (green line) gives rise to higher mass-loss, while an enhanced evaporation efficiency (violet line) also leads to higher mass-loss. A favourable combination of high XUV luminosity, low planetary bulk density, and high evaporation efficiency should give rise to maximal mass-loss. 

The additional effect of attenuation by the AGN dusty torus is further considered by \citet{Balbi_Tombesi_2017}. 
The XUV flux from the SMBH would then be reduced by a factor $\sim e^{-\tau}$, where $\tau$ is the optical depth of the obscuring torus ($\tau \sim 1$ in the equatorial plane and $\tau \sim 0$ in the polar directions). In this setting, rocky planets on $r \sim (0.1-1)$ kpc scales may eventually lose a mass comparable to the present-day Earth's atmospheric mass ($\Delta M_\mathrm{atm} \sim M_\mathrm{atm,\oplus}$), the exact amount depending on the different combinations of the AGN torus attenuation $\tau$ and the evaporation efficiency $\epsilon$ \citep{Balbi_Tombesi_2017}. In agreement with previous work, we conclude that the central SMBH radiation can potentially remove substantial amounts of atmospheric mass from exoplanets located in the Galactic bulge region. 

\section{ Observational constraints on the past activity history of Sgr A*}
\label{Sect_mass-loss_observational}

\begin{figure}
\begin{center}
\includegraphics[angle=0,width=0.4\textwidth]{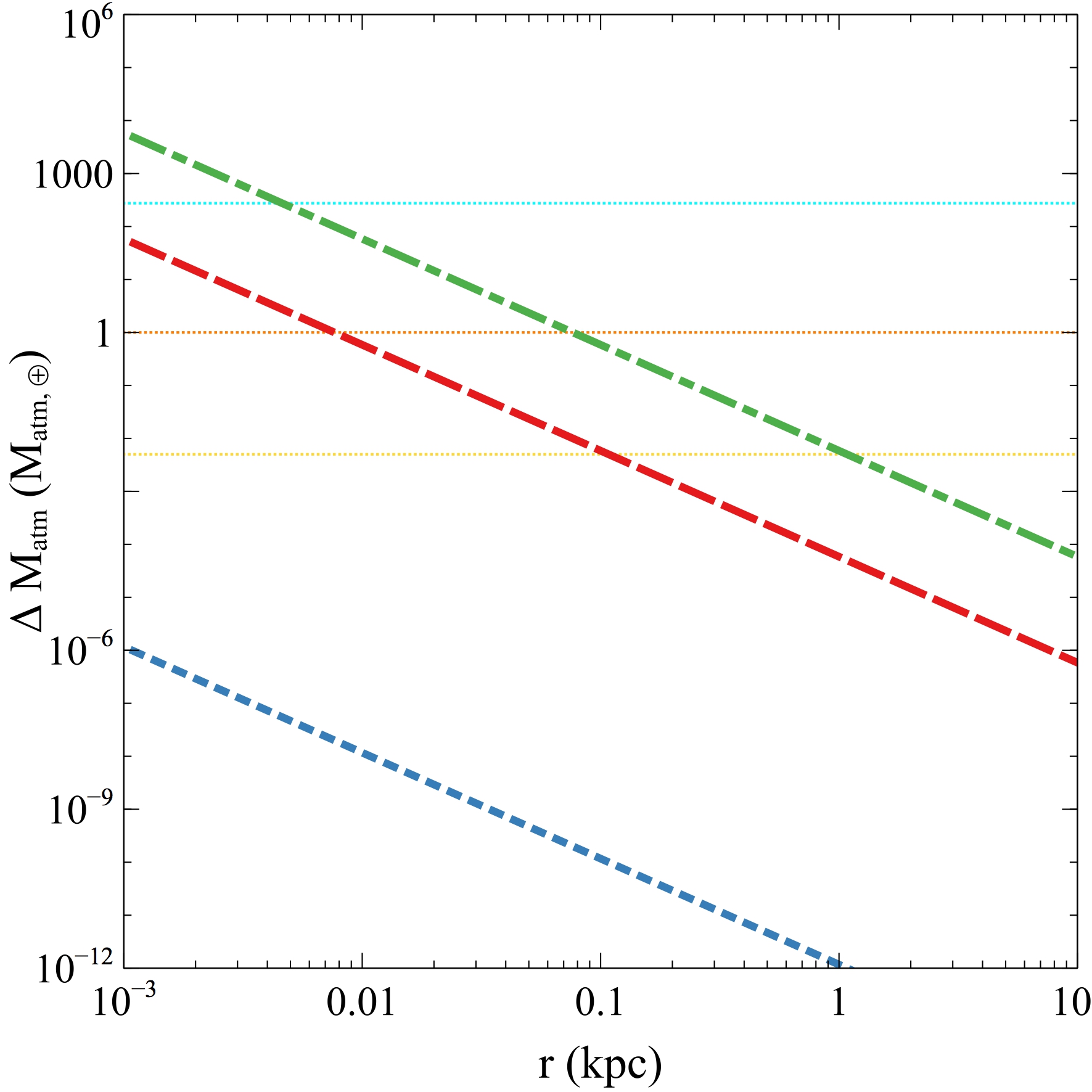}  
\caption{ 
Atmospheric mass-loss induced by the SMBH based on observational constraints of the past activity of Sgr A*. 
Variations in central luminosity and episode duration: an X-ray flare with $L = 10^{41}$ erg/s and $\Delta t = 1$ yr (blue dotted), a Seyfert-like episode with $L = 5 \times 10^{43}$ erg/s and $\Delta t = 10^5$ yr (red dashed), a quasar-like event with $L = 5 \times 10^{44}$ erg/s and $\Delta t = 10^6$ yr (green dash-dot). 
The three horizontal lines are the same as in Fig. \ref{Fig_DeltaM_r_model}.
}
\label{Fig_DeltaM_r_obs}
\end{center}
\end{figure} 

Currently, Sgr A* is in an extremely quiescent state, with a bolometric luminosity of just $L_\mathrm{bol} \lesssim 10^{36}$ erg/s, that is $\sim 9$ orders of magnitude below its Eddington limit \citep{EHT_2022_V}. So the present-day luminosity output of the central SMBH is insignificant and should have no appreciable effect on exoplanets within the Milky Way galaxy. 

However, there are several lines of empirical evidence indicating that Sgr A* was much brighter in the past, and that it previously underwent active episodes of gas accretion and associated energetic outbursts. The discovery of the Fermi bubbles, i.e. large gamma-ray bubbles extending $\sim 9$ kpc above and below the Galactic plane and symmetric with respect to the Galactic Centre, suggests a powerful energy injection event from the central SMBH sometime in the past \citep{Su_et_2010}. Relatedly, the recent identification of huge X-ray bubbles --called the eROSITA bubbles-- extending on $\sim 14$ kpc with a quasi-spherical configuration likely encompassing the Fermi bubbles, is again suggestive of a central SMBH outburst \citep{Predehl_et_2020}. 

The existence of such large-scale double symmetric structures emanating from the Galactic Centre --with similar morphologies in both $\gamma$-rays and X-rays-- points towards a common physical origin. Indeed, the Fermi and eROSITA bubbles, alongside the so-called microwave haze, have been interpreted as relics of past AGN activity originating from a previous energetic outburst of Sgr A* \citep{Yang_et_2022}. In order to account for the energy injection responsible for the inflation of the Fermi/eROSITA bubbles, the SMBH must have been accreting at significant fractions of the Eddington rate (with Eddington ratios $\lambda_\mathrm{E} \sim 0.1-1$). For instance, Sgr A* may have experienced an Eddington-limited AGN outburst event \citep{Zubovas_et_2011}, or a single jet activity episode \citep{Yang_et_2022}, occurring a few million years ago. So a Seyfert-like episode or even a quasar-like event (lasting $\sim 0.1 - 1$ Myr) may be required to simultaneously explain the Fermi and eROSITA bubbles. 

In addition, there are also indications of more recent nuclear activity at the Galactic Centre. X-ray polarisation measurements reveal short-lived X-ray flares from Sgr A*, which occurred just a few hundred ($\sim 100-200$) years ago \citep{Churazov_et_2017, Marin_et_2023}. These are likely responsible for the illumination of giant molecular clouds in the Galactic Centre region. The energetics of the X-ray flare is quite modest compared to the case of the Fermi/eROSITA bubbles, requiring much lower luminosities ($L \sim 10^{41}$ erg/s) and shorter flare durations (of a few years). 

All these observations support a picture of past AGN activity in our own Galactic Centre.  
Such empirical measures provide some constraints on the radiative output ($L$) as well as the episode duration ($\Delta t$). Based on the above discussion of the Fermi/eROSITA bubbles and X-ray flares, we consider three representative cases for the past activity history of Sgr A*: 
\begin{description}
\item[-] Quasar-like event: $L \approx 5 \times 10^{44}$ erg/s, $\Delta t \approx 10^6$ yr 
\item[-] Seyfert-like episode: $L \approx 5 \times 10^{43}$ erg/s, $\Delta t \approx 10^5$ yr 
\item[-] X-ray flare: $L \approx 10^{41}$ erg/s, $\Delta t \approx 1$ yr 
\end{description}

In Fig. \ref{Fig_DeltaM_r_obs}, we show the atmospheric mass-loss obtained from the observational constraints on Sgr A* history (which can be compared with the analogous plot in Fig. \ref{Fig_DeltaM_r_model}). We observe that the recent X-ray flare has almost no effect on the atmosphere of exoplanets: the mass-loss is negligible at all galactocentric radii, as the flare energetics is too low and the duration too short to have any meaningful impact (even in the innermost region). In contrast, a Seyfert-like episode can remove an atmospheric mass equivalent to Earth's atmosphere in the inner $\sim 10$ pc region, and also eject a Mars atmospheric mass out to a distance of $\sim 100$ pc. In the case of a powerful quasar-like event, an amount of mass comparable to Earth's ocean mass may be removed in the innermost zone, while a Martian atmospheric mass can be stripped on galactic scales ($\gtrsim 1$ kpc). Therefore the observational constraints confirm the notion that past Sgr A* activity might have considerably affected the evolution of exoplanets located in the inner region of the Galaxy. 


\section{Evolution of exoplanetary atmospheres under BH irradiation}
\label{Sect_atmospheric_evolution}


\subsection{Atmospheric mass fraction evolution}

We now examine the temporal evolution of the atmospheric mass fraction due to irradiation from the central SMBH. 
The atmospheric mass fraction of the exoplanet is defined as
\begin{equation}
X = \frac{M_\mathrm{atm}}{M_\mathrm{core}} , 
\end{equation}
where $M_\mathrm{atm}$ and $M_\mathrm{core}$ are the atmosphere mass and core mass, respectively. 
The evolution of the atmospheric mass fraction is described by a differential equation of the form 
\begin{equation}
\frac{dX}{dt} = - \frac{X}{t_\mathrm{\dot{X}}} , 
\label{Eq_X_tX}
\end{equation}
where $t_\mathrm{\dot{X}} = X/\dot{X} = M_\mathrm{atm}/\dot{M}_\mathrm{atm}$ is the atmospheric mass-loss timescale \citep{Rogers_Owen_2021}.  
Thus equation \ref{Eq_X_tX} can be recast as 
\begin{equation}
\frac{dX}{dt} 
= - \frac{\dot{M}_\mathrm{atm}}{M_\mathrm{core}} . 
\end{equation}
We recall that the mass-loss rate in the energy-limited formalism is given by $\dot{M}_\mathrm{atm} = \frac{3 \epsilon L_\mathrm{XUV}}{16 \pi G \rho_p r^2}$ (equation \ref{Eq_Mdot_Lxuv_bis}). 
Therefore the atmospheric evolution is governed by
\begin{equation}
\frac{dX}{dt} 
= - \frac{3 \epsilon L_\mathrm{XUV}(t)}{16 \pi G \rho_p r^2 M_\mathrm{core}} . 
\label{Eq_dX_dt}
\end{equation}
Integrating equation \ref{Eq_dX_dt}, we obtain the temporal evolution of the atmospheric mass fraction
\begin{equation}
X(t) = X_0 - \frac{3 \epsilon \int L_\mathrm{XUV}(t) dt }{16 \pi G \rho_p r^2 M_\mathrm{core}}  , 
\label{Eq_X_t}
\end{equation}
where $X_0$ is the initial atmospheric mass fraction. We note that in this framework, the exoplanet atmospheric evolution is directly tied to the XUV luminosity evolution and hence to the SMBH activity history. 

\begin{figure}
\begin{center}
\includegraphics[angle=0,width=0.4\textwidth]{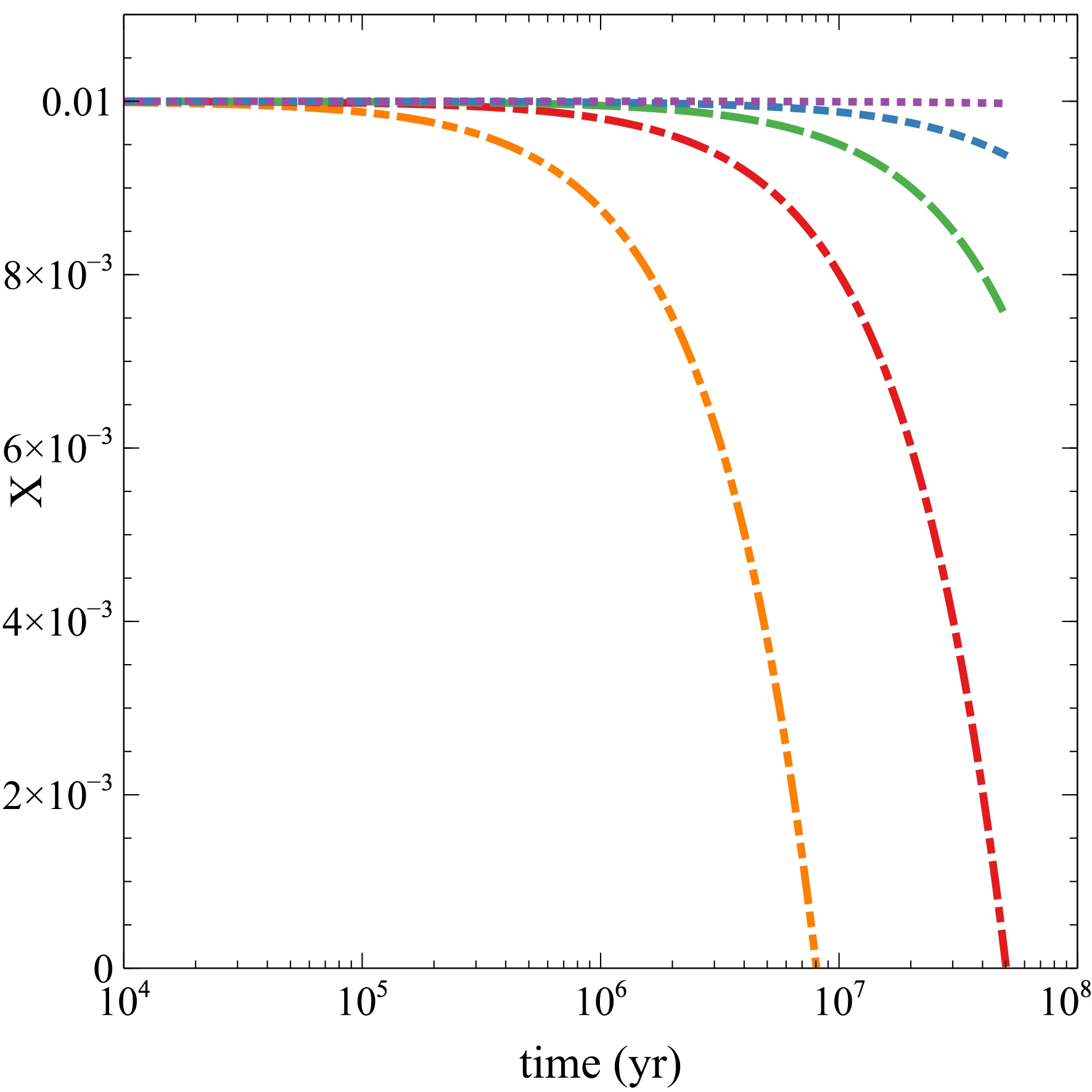} 
\caption{ 
Temporal evolution of the atmospheric mass fraction at different galactocentric radii: $r = 100$ pc (violet fine dotted), $r = 20$ pc (blue dotted), $r = 10$ pc (green dashed), $r = 5$ pc (red dash-dot), $r = 2$ pc (orange dash-dot-dot).
}
\label{Fig_X_t_var-r}
\end{center}
\end{figure}

Let us first consider the simple case of a constant luminosity output at the Eddington luminosity ($L_\mathrm{XUV}(t) = L_\mathrm{E} = 5 \times 10^{44}$ erg/s) lasting for a Salpeter time ($t = t_\mathrm{S}$). As a fiducial case, we consider an exoplanet with a terrestrial mass core ($M_\mathrm{core} = M_{\oplus}$) and bulk density $\rho_p = 5 \, \mathrm{g cm^{-3}}$, an initial atmospheric mass fraction of $X_0 = 0.01$, and a standard evaporation efficiency of $\epsilon = 0.1$. Figure \ref{Fig_X_t_var-r} shows the resulting atmospheric mass fraction as a function of time at different galactocentric distances. We see that at $r = 100$ pc, there is little evolution and the atmospheric mass fraction stays nearly constant, close to its initial value ($X(t) \sim X_0$). At a distance of $r = 10$ pc, an exoplanet is subject to non-negligible photoevaporation, and loses a significant fraction ($\sim 25 \%$) of its initial atmosphere. In the innermost regions ($r \lesssim 5$ pc), an exoplanet may be completely stripped of its initial atmosphere within the Salpeter time. A given planet can thus follow a completely different fate depending on its location in the Galaxy; naturally, the outcome will also depend on the AGN luminosity history and planetary parameters ---as analysed in the next subsections.

\subsection{SMBH luminosity temporal evolution}

\begin{figure*}
\begin{multicols}{2}
    \includegraphics[width=0.8\linewidth]{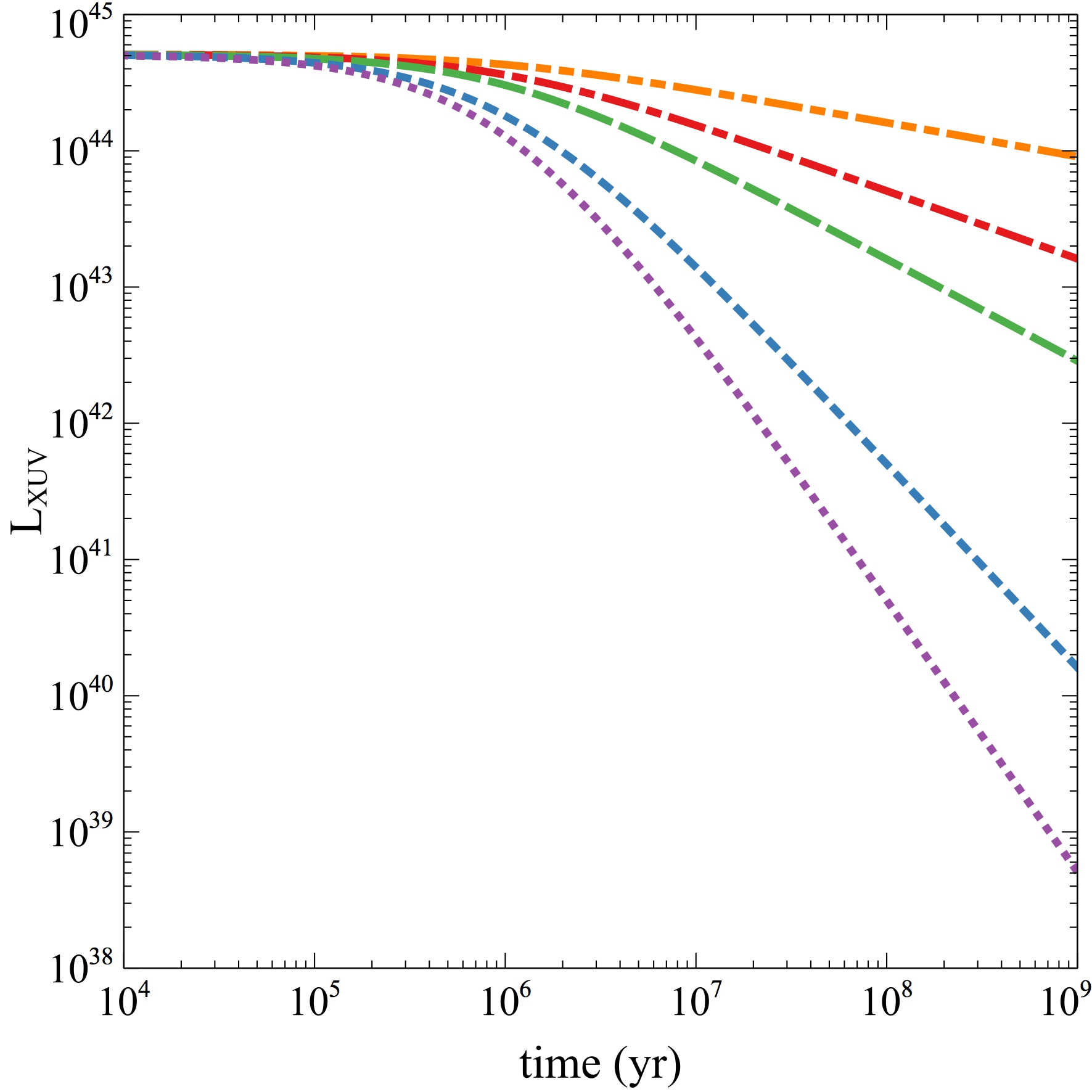}\par
    \includegraphics[width=0.8\linewidth]{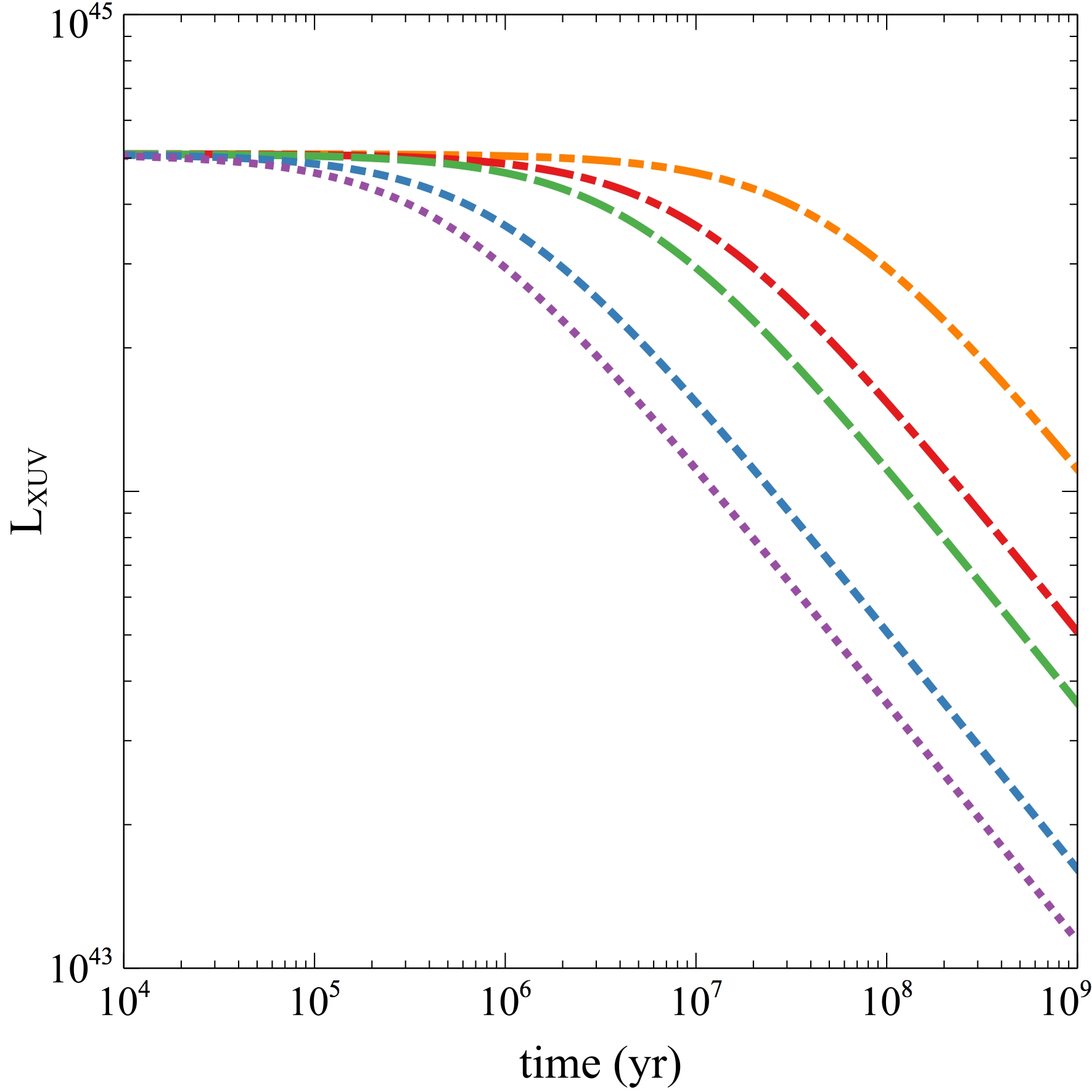}\par 
    \end{multicols}
\caption{ 
AGN luminosity temporal evolution following a power-law decay of the form $L_\mathrm{XUV,0} \left( 1 + t/t_d \right)^{-\delta}$, with $L_\mathrm{XUV,0} = 5.1 \times 10^{44}$ erg/s, $t_d = 10^6$ yr (left-hand panel), $\delta = 0.5$ (right-hand panel). Variations in power-law slope (left-hand panel): $\delta = 2$ (violet fine dotted), $\delta = 1.5$ (blue dotted), $\delta = 0.75$ (green dashed), $\delta = 0.5$ (red dash-dot), $\delta = 0.25$ (orange dash-dot-dot). Variations in AGN episode duration (right-hand panel): $t_d = 5 \times 10^5$ yr (violet fine dotted), $t_d = 10^6$ yr (blue dotted), $t_d = 5 \times 10^6$ yr (green dashed), $t_d = 10^7$ yr (red dash-dot), $t_d = 5 \times 10^7$ yr (orange dash-dot-dot).
}
\label{Fig_Lxuv_t}
\end{figure*}

\begin{figure*}
\begin{multicols}{2}
    \includegraphics[width=0.8\linewidth]{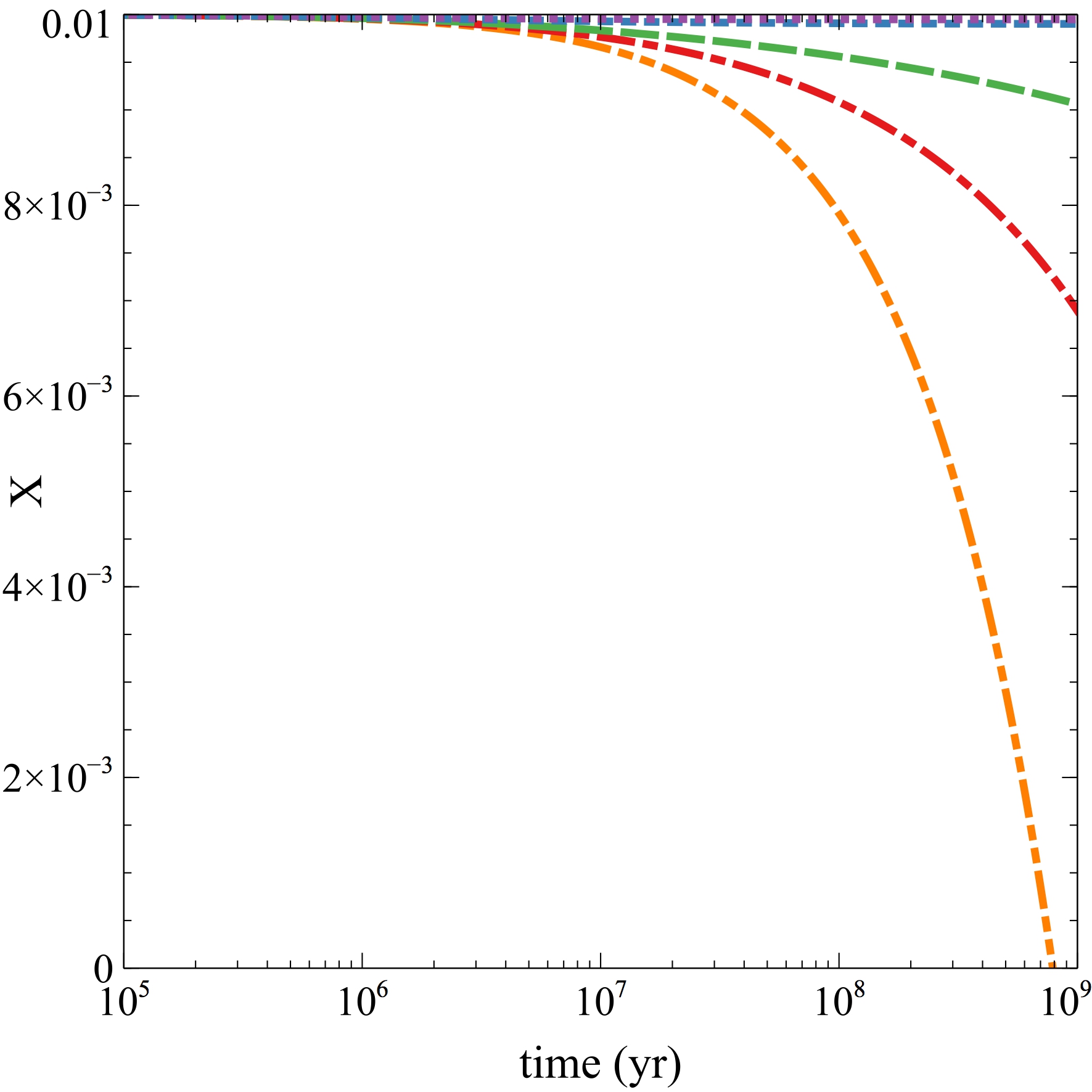}\par
    \includegraphics[width=0.8\linewidth]{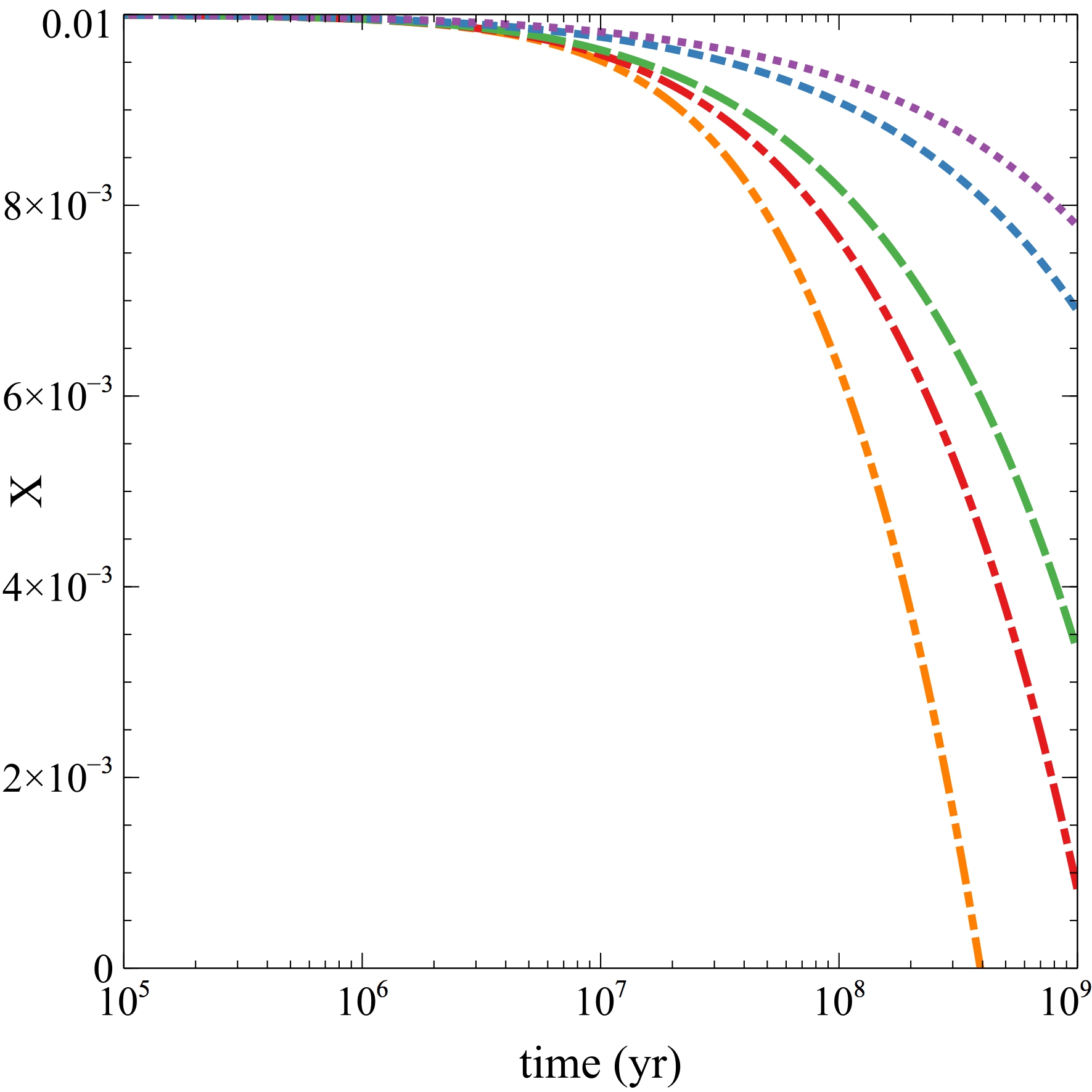}\par 
    \end{multicols}
\caption{ 
Temporal evolution of the atmospheric mass fraction corresponding to the AGN luminosity evolution plotted in Fig. \ref{Fig_Lxuv_t}. Fiducial exoplanet parameters: $\epsilon = 0.1$, $\rho_p = 5.0 \, \mathrm{g cm^{-3}}$,  $M_\mathrm{core} = M_\mathrm{\oplus}$, $X_0 = 0.01$. Variations in power-law slope (left-hand panel): $\delta = 2$ (violet fine dotted), $\delta = 1.5$ (blue dotted), $\delta = 0.75$ (green dashed), $\delta = 0.5$ (red dash-dot), $\delta = 0.25$ (orange dash-dot-dot). Variations in AGN episode duration (right-hand panel): $t_d = 5 \times 10^5$ yr (violet fine dotted), $t_d = 10^6$ yr (blue dotted), $t_d = 5 \times 10^6$ yr (green dashed), $t_d = 10^7$ yr (red dash-dot), $t_d = 5 \times 10^7$ yr (orange dash-dot-dot). 
}
\label{Fig_X_t_var-L}
\end{figure*}

In general, previous works have only considered a constant luminosity output from the central SMBH lasting for a Salpeter time \citep[e.g.][]{Forbes_Loeb_2018, Wislocka_et_2019}. 
In more realistic situations, the AGN radiative output does not stay constant over time, and is likely to decline following an accretion event. Different forms of AGN luminosity decay can be considered, such as exponential fall-offs or power-law decays  \citep{Ishibashi_Fabian_2018}. If the accretion disc slowly dissipates on a viscous timescale following an accretion episode, the resulting AGN luminosity is more likely to follow a power-law decay \citep{Zubovas_2018}.  
The temporal evolution of the BH luminosity can be described by a power-law of the form
\begin{equation}
L_\mathrm{XUV}(t) = L_\mathrm{XUV,0} \left( 1 + \frac{t}{t_d} \right)^{-\delta} , 
\label{Eq_Lxuv_PL}
\end{equation} 
where $L_\mathrm{XUV,0}$ is the initial XUV luminosity, $\delta$ is the power-law slope, and $t_d$ is the characteristic timescale of AGN episode duration. 

Figure \ref{Fig_Lxuv_t} shows the temporal evolution of the AGN luminosity output, for different power-law slopes $\delta$ (left-hand panel) and episode duration $t_d$ (right-hand panel). Here we focus on exoplanets located in the galactic nucleus region, with $r = 10$ pc as a fiducial value. We see that at early times (in the limit $t/t_d \ll 1$), the luminosity is roughly constant $L_\mathrm{XUV}(t) \sim L_\mathrm{XUV,0}$; it subsequently declines as $L_\mathrm{XUV}(t) \sim L_\mathrm{XUV,0} (t/t_d)^{-\delta}$ at late times ($t/t_d \gg 1$). Variations in the power-law slope can cause strong luminosity evolution, with steeper slopes leading to luminosity fall-off by several orders of magnitude; while variations in AGN episode duration generate milder luminosity decay (at least for the considered range of $t_d$). 

We can now evaluate how the AGN luminosity history affects the exoplanet atmospheric evolution. Integrating the radiative luminosity output (equation \ref{Eq_Lxuv_PL}), we obtain
\begin{equation}
\int L_\mathrm{XUV}(t) dt 
= \frac{L_\mathrm{XUV,0} t_d}{1-\delta} \left[ \left( 1 + \frac{t}{t_d} \right)^{1-\delta} - 1 \right] . 
\end{equation} 
As a result, for a power-law luminosity decay, the atmospheric mass fraction (equation \ref{Eq_X_t}) evolves as 
\begin{equation}
X(t) = X_0 - \frac{3 \epsilon L_\mathrm{XUV,0} t_d}{(1-\delta)16 \pi G \rho_p r^2 M_\mathrm{core}} \left[ \left( 1 + \frac{t}{t_d} \right)^{1-\delta} - 1 \right] . 
\end{equation}

In Fig. \ref{Fig_X_t_var-L}, we show the corresponding temporal evolution of the atmospheric mass fraction matching the AGN luminosity evolution plotted in Fig. \ref{Fig_Lxuv_t}. We observe that the AGN luminosity history has a major impact on the evolution of the planetary atmosphere: the gaseous envelope can either survive unscathed or be completely lost --depending on the BH activity history. For steep power-law slopes ($\delta > 1$), the central luminosity rapidly decreases, and there is negligible atmospheric mass-loss. On the other hand, the luminosity decay is slower for shallower power-law slopes and longer episode durations, hence the atmospheric mass-loss becomes more important. In some cases (with shallow slope $\delta \lesssim 0.25$ and long duration $t_d \gtrsim 10^7$ yr), the AGN stays relatively luminous for a prolonged time, such that the atmospheric mass may be completely removed. 


\subsection{Dependence on exoplanetary parameters}

\begin{figure*}
\begin{multicols}{3}
       \includegraphics[width=\linewidth]{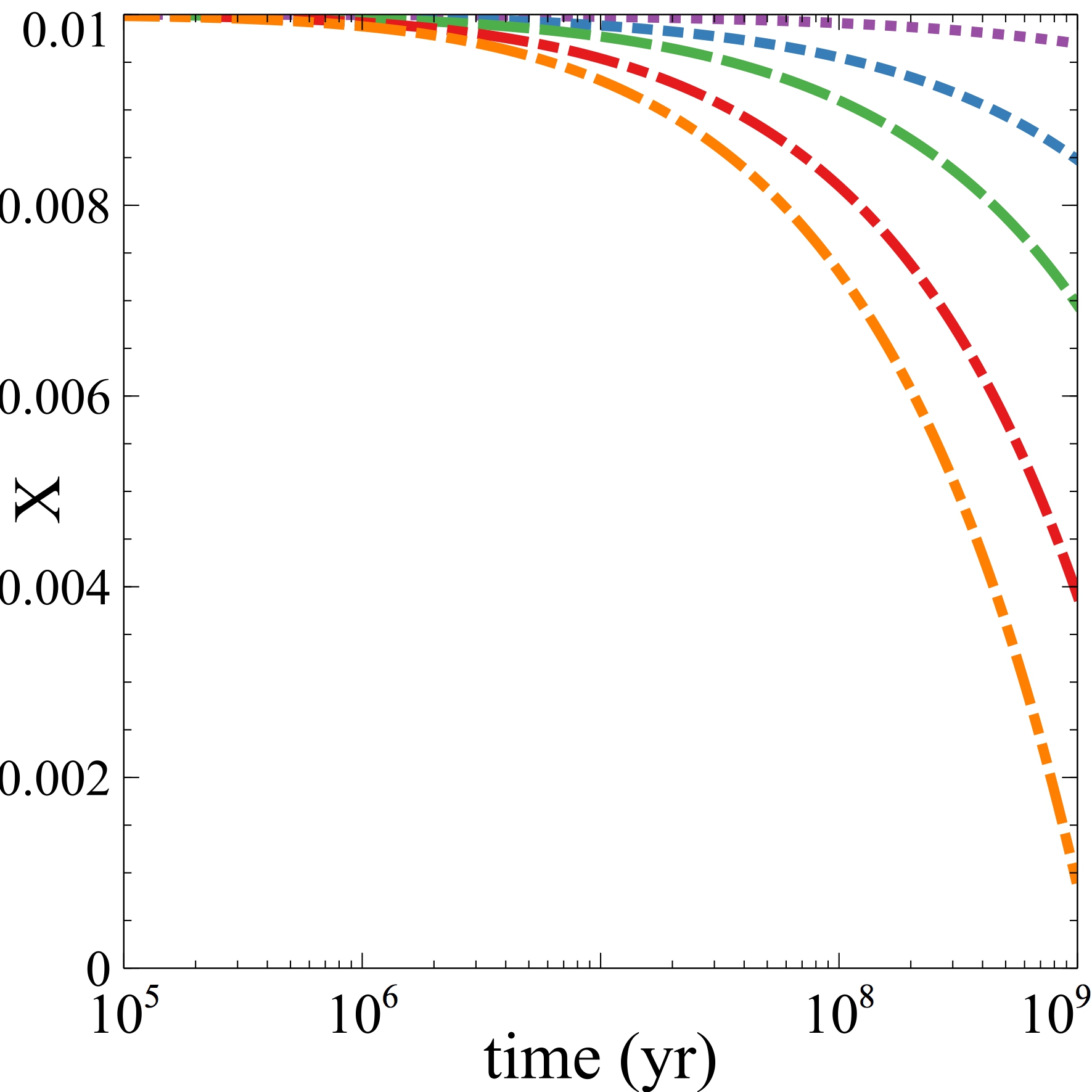}\par
       \includegraphics[width=\linewidth]{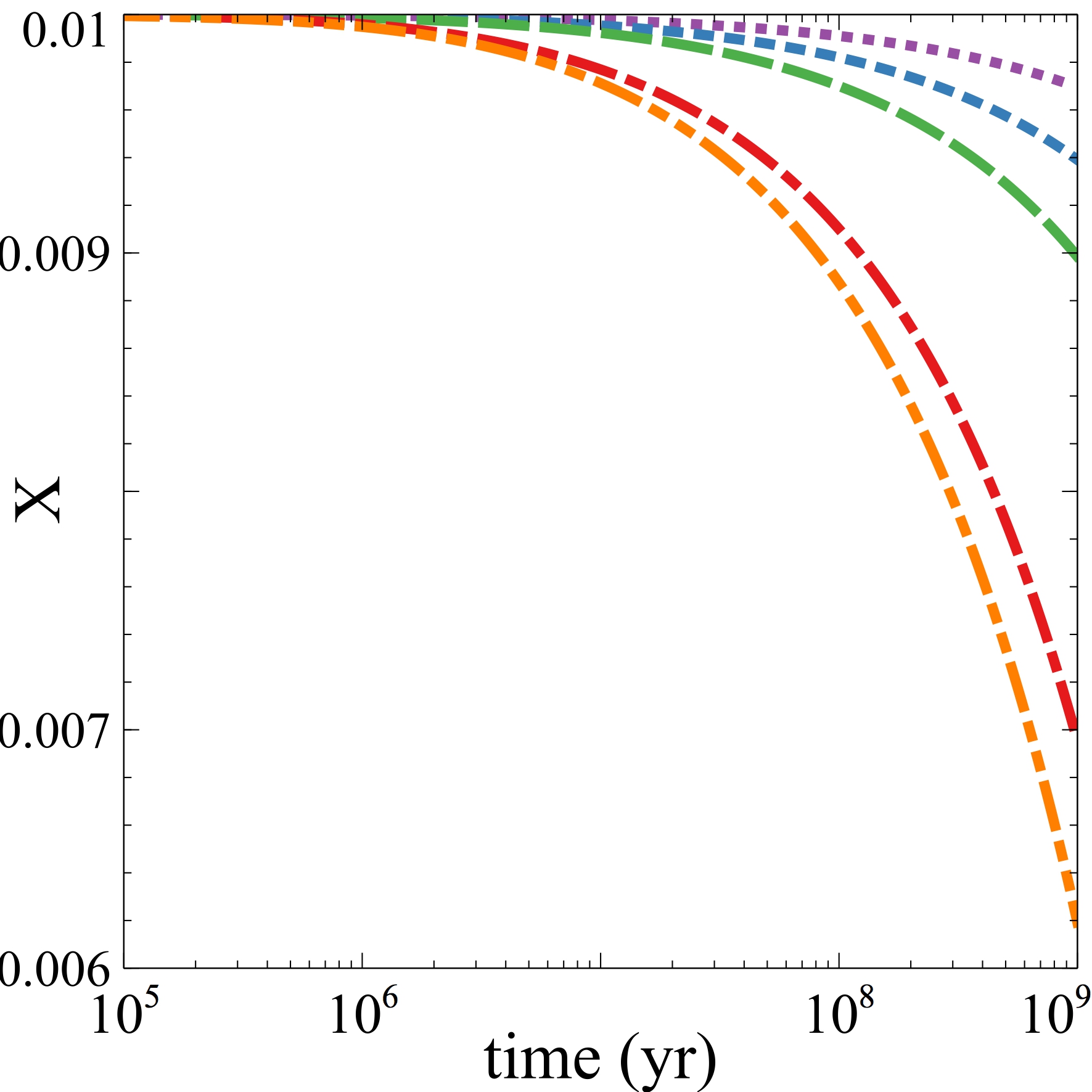}\par 
       \includegraphics[width=\linewidth]{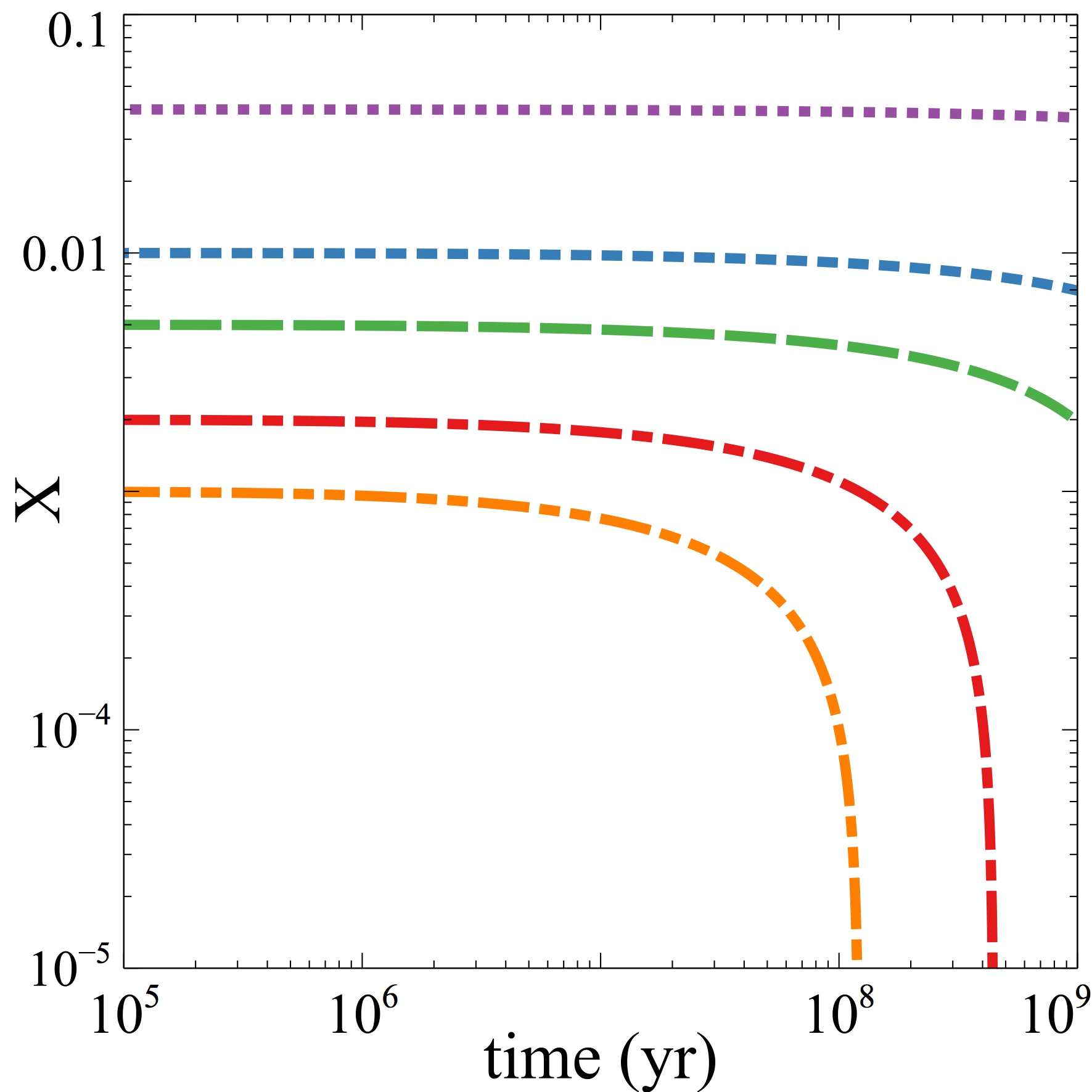}\par 
 \end{multicols}
\caption{ 
Temporal evolution of the atmospheric mass fraction with fiducial values ($\epsilon = 0.1$, $\rho_p = 5 \, \mathrm{g cm^{-3}}$, $M_\mathrm{core} = M_{\oplus}$, $X_0 = 0.01$) and variations in the exoplanet parameters. 
Variations in evaporation efficiency (left-hand panel): $\epsilon = 0.01$ (violet fine dotted), $\epsilon = 0.05$ (blue dotted), $\epsilon = 0.1$ (green dashed), $\epsilon = 0.2$ (red dash-dot), $\epsilon = 0.3$ (orange dash-dot-dot).
Variations in core mass (middle panel): $M_\mathrm{core} = 10 M_\mathrm{\oplus}$ (violet fine dotted), $M_\mathrm{core} = 5 M_\mathrm{\oplus}$ (blue dotted), $M_\mathrm{core} = 3 M_\mathrm{\oplus}$ (green dashed), $M_\mathrm{core} = 1 M_\mathrm{\oplus}$ (red dash-dot), $M_\mathrm{core} = 0.8 M_\mathrm{\oplus}$ (orange dash-dot-dot). Variations in initial atmospheric mass fraction (right-hand panel): $X_0 = 0.04$ (violet fine dotted), $X_0 = 0.01$ (blue dotted), $X_0 = 0.005$ (green dashed), $X_0 = 0.002$ (red dash-dot), $X_0 = 0.001$ (orange dash-dot-dot).
}
\label{Fig_X_t_var-exp}
\end{figure*}

We next consider variations in the physical parameters of the exoplanets: the evaporation efficiency ($\epsilon$), the core mass ($M_\mathrm{core}$), and the initial mass fraction ($X_0$). Here we assume a typical AGN luminosity decay history characterised by $L_\mathrm{XUV,0}= L_\mathrm{E}$, $\delta = 0.5$ and $t_d = 10^6$ yr. 

Figure \ref{Fig_X_t_var-exp} shows the temporal evolution of the atmospheric mass fraction for different evaporation efficiencies (left-hand panel), core masses (middle panel), and initial mass fractions (right-hand panel). We see that the evaporation efficiency can have a significant effect on the atmospheric evolution: there is almost no mass-loss for $\epsilon = 0.01$, whereas the atmosphere is close to being completely stripped for $\epsilon = 0.3$. The core mass can also play an important role in atmospheric retention. For a core mass comparable or smaller than Earth mass ($M_\mathrm{core} \lesssim M_\mathrm{\oplus}$), a planet undergoes significant mass-loss with $\gtrsim 30 \%$ of the initial atmosphere being removed. The photoevaporation is less effective for higher core masses, e.g. for $M_\mathrm{core} \geq 3 M_\mathrm{\oplus}$, the total mass-loss is less than $\sim 10 \%$. Concerning the initial atmospheric mass fraction, we note that there is little mass-loss for $X_0 \sim 4 \%$, whereas a planet can be completely stripped of its initial atmosphere if $X_0 \lesssim 0.2 \%$. 


\section{ The case of stellar-mass BH}
\label{Sect_smBH}

Up to now, we have focused on the central SMBH, Sgr A*. But there ought to be millions of stellar-mass BHs --as remnants of massive star evolution-- distributed throughout the Galaxy. In particular, a high concentration of stellar-mass BHs (a density cusp) is observed in the Galactic Centre region \citep{Hailey_et_2018}. Stellar-mass BHs, with a typical mass of $M_\mathrm{BH} \sim 10 M_{\odot}$, accrete matter from the companion star in binary systems, and are usually observed as X-ray binaries \citep{Remillard_McClintock_2006}. Accreting stellar-mass BHs emit high-energy radiation into the surrounding environment, hence may play an analogous role to Sgr A*, albeit on reduced local scales. 

To zeroth order, the accretion disc spectrum can be modelled by blackbody radiation, $\sigma T^4(R)$, where $\sigma$ is the Stefan-Boltzmann constant and $T(R)$ is the blackbody temperature at radius $R$. The temperature at the inner radius of the disc is given by \citep{Pringle_1981}
\begin{equation}
T_\mathrm{in} \sim \left( \frac{3 G M \dot{M}}{8 \pi \sigma R_\mathrm{in}^3} \right)^{1/4} , 
\end{equation}
which roughly defines the location of the peak in the emitted spectrum, where most of the radiative power is actually emitted. The peak value is reached at radius $R_\mathrm{in} = \frac{25}{9} R_\mathrm{ISCO}$ \citep{Reynolds_2021}, where $R_\mathrm{ISCO}$ is the innermost stable circular orbit (ISCO). For a non-rotating Schwarzschild BH, the ISCO is located at $R_\mathrm{ISCO} = 3 R_\mathrm{S} = 6 R_\mathrm{G}$, where $R_\mathrm{G} = G M_\mathrm{BH} / c^2$ is the gravitational radius. Introducing the Eddington ratio $\lambda_\mathrm{E} = L/L_\mathrm{E} = \dot{M}/\dot{M}_\mathrm{E}$, where $\dot{M}_\mathrm{E} = L_\mathrm{E}/\eta c^2$ and $\eta = 0.1$ is the radiative efficiency, the peak temperature is given by 
\begin{equation}
T_\mathrm{in} = \left( 3.2 \times 10^{-4} \frac{c^5 m_p \lambda_\mathrm{E}}{\eta \sigma \sigma_T G M_\mathrm{BH}} \right)^{1/4} \mathrm{K} . 
\end{equation}

Numerically, for a stellar-mass BH of ten solar masses ($M_\mathrm{BH} = 10 M_{\odot}$) accreting at ten percent of the Eddington rate ($\lambda_\mathrm{E} = 0.1$), the characteristic temperature is $T_\mathrm{in} \sim 4 \times 10^6$ K, corresponding to an energy of $E \sim k_\mathrm{B} T \sim 0.3$ keV, with a peak in the soft X-rays. 

The blackbody luminosity is defined as
\begin{equation}
L = 4 \pi R_\mathrm{in}^2 \sigma T_\mathrm{in}^4 , 
\end{equation}
and the corresponding X-ray luminosity of the stellar-mass BH is about $L_\mathrm{X} \sim 1.1 \times 10^{38}$ erg/s.  
This is several orders of magnitude higher than the typical X-ray luminosities ($L_\mathrm{X} \sim 10^{28} - 10^{30}$ erg/s) of young stars in open clusters \citep{Ketzer_et_2024}. 

Assuming that most of the high-energy radiation contributes to drive atmospheric escape, one can estimate the maximal distance where the X-ray luminosity exceeds the minimum flux required to drive a hydrodynamic outflow: $r_\mathrm{max} =\sqrt{ \frac{L_\mathrm{X}}{4 \pi F_\mathrm{min}}}$, which is about $r_\mathrm{max} \sim 3.1$ pc (for $F_\mathrm{min} = 0.1 \, \mathrm{erg s^{-1} cm^{-2}}$, cf. Sect. \ref{Sect_mass-loss_theoretical}). Thus the maximal distance is around a few parsecs for the fiducial case. In the case of heavier stellar-mass BHs ($M_\mathrm{BH} \sim 20-30 M_{\odot}$), such as Cygnus X-1 \citep{Miller-Jones_et_2021} or the recently discovered Gaia BH3 \citep{GAIA_et_2024}, and/or sources accreting at higher rates close to the Eddington limit ($L/L_\mathrm{E} \sim 1$), the influence distance can be greater, of the order of $r_\mathrm{max} \gtrsim 10$ pc. This suggests that stellar-mass BHs may also have a significant impact on the evolution of exoplanets in the local galactic surroundings. 


\section{ Discussion }
\label{Sect_Discussion}

\subsection{Implications for exoplanetary habitability}

The habitable zone is defined as the region around a host star where liquid water can exist and be maintained on the surface of an exoplanet. While atmospheres may be necessary for sustaining life, a thick H/He envelope generates high surface pressure and temperature that would prevent the existence of liquid water on the surface and thus preclude habitability. As a consequence, if exoplanets retain their voluminous H/He envelopes throughout their lifetime, they would not be habitable ---at least in the conventional sense \citep{Owen_Mohanty_2016}. 

In this context, XUV photoevaporation can provide a means to get rid of the primordial gaseous envelope, making the planet potentially habitable \citep{Luger_et_2015, Owen_Mohanty_2016, Chen_et_2018}. If a large fraction of the atmospheric mass can be removed ($ X \ll 10^{-3}$) --leading to a significant reduction in the surface temperature and pressure-- favourable conditions for habitability may then be achieved. It has even been argued that XUV photoevaporation could help transform gaseous sub-Neptunes into rocky super-Earths that are potentially habitable, the so-called `habitable evaporated cores' \citep{Luger_et_2015}. 

Here we have shown that XUV irradiation from the SMBH can induce significant atmospheric escape, with the final outcome depending on the BH accretion history and the nature of the exoplanet itself (Sect. \ref{Sect_atmospheric_evolution}). In some cases, the initial atmosphere may be stripped away for an exoplanet located in the Galactic Centre region. This could possibly support the emergence of bare terrestrial cores from the erosion of gas-rich sub-Neptunes in the vicinity of Sgr A* \citep{Chen_et_2018}. Other studies further speculate that AGN radiation may induce additional beneficial effects, such as stimulated prebiotic chemistry and photosynthesis on free-floating planets \citep{Lingam_et_2019}. In this perspective, the central BH may play a positive role in promoting planetary habitability and ultimately life in the Universe.

On the other hand, it is well known that a partial or complete disruption of the planetary atmosphere will have a negative impact on surface life. In the absence of a protective shield, some of the high-energy ionising photons may penetrate deep into the atmosphere and cause severe biological damage. For instance, the UV flux from Sgr A* may exceed the lethal dose for putative microorganisms living on nearby planets \citep{Balbi_Tombesi_2017}. Furthermore, X-rays could trigger the depletion of atmospheric ozone in catalytic cycles \citep{Ambrifi_et_2022}. Hard X-ray photons can break the nitrogen bond ($\mathrm{N_2}$) and generate nitrogen oxides ($\mathrm{NO_x}$), which in turn leads to the destruction of ozone ($\mathrm{O_3}$). It has been estimated that, for an Earth-like planet, an average ozone depletion of $\sim 30 \%$ could lead to a mass extinction event \citep[][and references therein]{Brunton_et_2023, Perkins_et_2024}. 

The actual outcome will also depend on the development of a secondary atmosphere (e.g. by volcanic outgassing) that could mitigate some of the harmful radiation effects. The negative effects associated with the central SMBH would limit the viability of life-hosting planets in the Galactic bulge, with implications for the Galactic Habitable Zone \citep[][and references therein]{Balbi_et_2020}. Overall, there is a complex interplay between BH radiation, atmospheric chemistry, and biological evolution. More generally, the evolution of exoplanets located in the Galactic bulge region ($\lesssim 1$ kpc) may be significantly affected by the presence of the central SMBH and its accretion history over cosmic time ---with both positive and negative ramifications. 


\subsection{ Some caveats and future outlook }

Here we have considered the standard energy-limited formalism, with a constant mass-loss efficiency ($\epsilon = 0.1$), whereby the mass-loss rate scales linearly with the incident flux ($\dot{M} \propto F_\mathrm{XUV}$). Although the energy-limited prescription is widely adopted in the literature, it suffers from a number of known limitations \citep[][]{Owen_2019, Krenn_et_2021}. Actually, the efficiency parameter depends on the planet properties (mass and radius) and the incident high-energy flux. At high XUV fluxes, the efficiency declines and the flow enters the recombination-limited regime \citep{Murray-Clay_et_2009}; whereas at low XUV fluxes, the efficiency drops --as the hydrodynamic limit is no longer applicable-- and the flow transitions into Jeans escape  characterised by reduced mass-loss rate \citep{Owen_Mohanty_2016}.

As a consequence, the energy-limited approximation tends to over-estimate the mass-loss rate, although there are other instances where it can lead to under-estimates \citep{Kubyshkina_et_2018, Krenn_et_2021}. A more accurate treatment of the atmospheric escape process can only be obtained through numerical hydrodynamical simulations \citep[][and references therein]{Owen_2019}. An example is the hydro-approximation, based on a grid of one-dimensional hydrodynamic atmosphere models covering a wide range of exoplanet parameters \citep{Kubyshkina_et_2018}. In principle, reversing the argument, one might be able to infer the initial planetary conditions, alongside the past BH activity history, from the observable properties of present-day exoplanets.

Concerning BH radiation, the modelling aspect can be improved by considering more realistic spectral energy distributions. The location of the peak in the emitted spectrum depends on the BH intrinsic properties (mass and spin), as well as the accretion flow mode. The accretion disc spectrum is shifted towards higher energies for lower BH mass: the peak emission is located in the UV region for SMBH like Sgr A*, while the emission peaks in the X-rays for stellar-mass BHs. Similarly, high BH spins will lead to harder spectra in rapidly rotating sources \citep{Reynolds_2021}. The exact radiation pattern, including the BH mass and spin dependence, will need to be computed via numerical codes including relativistic effects \citep{Ishibashi_Fabian_Reynolds_2019}. Furthermore, the overall shape of the spectral energy distribution is governed by changes in the underlying accretion flow mode ---either radiatively efficient accretion discs or radiatively inefficient accretion flows. The resulting differences in the dominant peak location (UV or X-rays) may have distinct effects on the exoplanet atmospheric escape and potential biological hazard. 

Unlike previous works that only examined the case of a constant radiative output from Sgr A*, here we explicitly consider the temporal evolution of the SMBH luminosity, $L(t)$. Assuming different forms of AGN luminosity decay, we analyse its effects on the atmospheric evolution of surrounding exoplanets (Sect. \ref{Sect_atmospheric_evolution}). On a related note, tidal disruption events (TDE) may provide another source of time-variable high-energy radiation, which could also affect the exoplanets and their habitability \citep{Pacetti_et_2020}. In the case of tidal disruption of stars approaching the central SMBH, the time evolution of the accretion rate can be modelled as a power-law, following a typical decline of the form $\propto t^{-5/3}$. \citet{Pacetti_et_2020} analyse the cumulative effect of such tidal disruption events in the Galaxy, showing that TDE can induce an atmospheric mass loss comparable to AGN radiation on galactic scales ($r \sim 0.1-1$ kpc). 

Another interesting aspect is the angular dependence of the AGN radiative output, $L(\theta)$, which should affect the spatial distribution of exoplanets in the Milky Way galaxy \citep[see also][]{Balbi_Tombesi_2017}. 
Such directional dependence could arise from the relative orientations of the central BH spin axis, the AGN dusty torus, and the stellar disc on large scales. This implies that exoplanets located in different regions of the Galaxy --galactic plane vs. polar directions-- may follow diverse fates. Naturally, this will have important implications for galactic habitability and may require a revisit of the Galactic Habitable Zone. While we have focused on the effects of SMBH radiation, AGN-driven winds and galactic outflows might also play a role in the atmospheric evolution of exoplanets \citep{Ambrifi_et_2022, Heinz_2022}. We conclude that accreting BHs may have a non-negligible impact on the evolution of exoplanets in our Galaxy, and their multiple roles --including both positive and negative effects-- deserve to be further investigated. 


\section*{Data availability}

No new data were generated or analysed in support of this research.


\bibliographystyle{mn2e}
\bibliography{biblio.bib}

\label{lastpage}

\end{document}